
%
\magnification=\magstep1
%

\font\sc=cmcsc10
\def\ref#1\par{\parshape=2 0.0truecm 15.2truecm 1.5truecm 14.2truecm {#1}
\par}
\def\kms{km s$^{\rm -1}$\ }

\def\oneskip{\vskip\baselineskip}
\def\etal{{\it et al.}\ }
\def\eg{{\it e.g.,}\ }
\def\CAII{{Ca~{\sc ii}}\ }
\def\hmed{$\langle{\rm H}_{\beta\gamma\delta}\rangle$\ }
\def\gsim{\ifmmode{\mathrel{\mathpalette\@versim>}}
    \else{$\mathrel{\mathpalette\@versim>}$}\fi}
\def\lsim{\ifmmode{\mathrel{\mathpalette\@versim<}}
    \else{$\mathrel{\mathpalette\@versim<}$}\fi}
\def\@versim#1#2{\lower 2.9truept \vbox{\baselineskip 0pt \lineskip
    0.5truept \ialign{$\m@th#1\hfil##\hfil$\crcr#2\crcr\sim\crcr}}}
%
\baselineskip=20pt
\overfullrule=0pt
\raggedbottom
\null
\oneskip
\oneskip
\centerline{\bf Spectroscopy
of Dwarf Elliptical Galaxies}
\centerline{\bf in the Fornax Cluster\footnote{$^1$}
{\rm Based on data obtained at the European Southern Observatory,
 La Silla, Chile}
}
\oneskip
\centerline {Enrico V. Held}
\centerline {Osservatorio Astronomico di Bologna --
             via Zamboni 33, Bologna 40126, Italy}
\centerline {and}
\centerline {Jeremy R. Mould}
\centerline {Palomar Observatory, California Institute of Technology --
             Pasadena, California  91125, U.S.A.}
\vfill
\noindent
      The Astronomical Journal, in press
\oneskip
\noindent
\break

\centerline{ABSTRACT}

We present the results of spectroscopic observations of 10 nucleated
dwarf elliptical galaxies (dE's) in the Fornax cluster. The blue spectra
of Fornax dE galaxies  indicate a wide range of metallicities at a given
luminosity, similar to those of {\it intermediate} to {\it metal-rich}
globular clusters. Metal abundances derived in this paper are well
correlated with optical colors and agree with previous spectroscopic
results. A discrepancy with metallicities inferred from infrared colors
is evident; possible causes include an intermediate age population and
dilution of spectral features by a blue light excess. Dwarf ellipticals
exhibit a wide variation of hydrogen line strength which points to a
complex star formation history. Prominent Balmer absorption lines are
the signature of a young stellar population in the nuclei of some (but
not all) dE's, while moderately strong Balmer lines in relatively
metal-rich dE's are more consistent with an extended main sequence. In a
few metal-poor dE galaxies, the hydrogen lines are consistent with or
perhaps weaker than those found in Galactic globulars of similar
metallicity. In the limited magnitude range of this sample, there is no
apparent correlation of metallicity either with effective and central
surface brightness, or with total and nuclear magnitudes. The velocity
distribution of the Fornax dwarfs is flatter than that of brighter
galaxies at the 75\% confidence level, possibly indicating a difference
in the kinematics of the two samples.

\vfill
\noindent
{\it Subject headings:} clusters: galaxies -- galaxies: kinematics --
galaxies: stellar content
\vfill\eject

\centerline{1. INTRODUCTION}
\oneskip
\noindent

Noting that {\it the star formation histories of dwarf spheroidal
galaxies are clearly complex and varied}, Da Costa (1988) remarks in
his review that the histories of
more luminous dwarf ellipticals (dE's) are also likely to
be complex and varied.
Indeed, broad band photometry has provided hints of complex star
formation histories in dwarf ellipticals (Caldwell 1983,
Caldwell \& Bothun 1987, hereafter CB87).
These authors showed that dwarf ellipticals have bluer $U-B$ colors than
globular clusters with similar $B-V$, which may indicate an additional
contribution to the B light. They also argued that the dispersion in
$U - V$ colors is too large to be explained by a pure metallicity
variation.
In a few cases, dwarf ellipticals are bluer in the $B - H$
optical-infrared colors than globular clusters
of similar $J - K$ color (Bothun \& Caldwell 1984, Zinnecker \etal
1985), a result which may indicate the presence of a warmer stellar
component, possibly a younger main-sequence turn-off.
In general, $B - H$ colors of dE's, which measure the relative
importance of giant and main-sequence populations, are typical of
metal-rich globular clusters (Bothun \& Caldwell 1984) -- the exact
range is made, however, somewhat uncertain by the use of photographic
blue magnitudes.

Additional evidence for an intermediate age population came from
spectroscopy (Bothun \etal 1985, Bothun \& Mould 1988).
In a study of 12 Virgo dE's, Bothun \& Mould (1988) found that dE spectra
resemble those of metal-rich globular clusters (GC's), but with Balmer lines
most likely due to an extended (F-type) main sequence. Also, in a recent
brief paper, Gregg's (1992) argues that nucleated dwarfs have too small
a 4000 \AA\ break for their Mg strength, possibly due to an intermediate age
population.

Since in evolved populations, the UV-blue spectral region is dominated
by light from the main-sequence turn-off, blue spectroscopy allows one
to study directly the young and the metal-poor population, while
infrared photometry samples the red giants. In this paper we have
obtained spectroscopy of a sample of 10 dwarf ellipticals in the Fornax
cluster. New measurements of metal and hydrogen lines have been combined
with available photometric data to discuss the implications for the
stellar content of dE's. Although our conclusions will strictly apply to
nuclei, CCD photometry indicates no difference in color between the
nuclei and the body of the galaxies  (CB87).


Another aim of this paper is to use our measurements of radial
velocity to investigate the velocity distribution of dwarf ellipticals
in the Fornax cluster together with that of brighter cluster members.
In the Nearby Galaxy Catalog (Tully 1988) the Fornax and Virgo clusters
and the IC 342 group have the highest galaxy volume density in the Local
Supercluster. The IC 342 group is a local group of modest total
luminosity, but Fornax has more than twice the central surface density
of the Virgo cluster (Ferguson \& Sandage 1988). It represents a unique
environment for understanding the formation and dynamics of clusters of
galaxies.

Recent important contributions to the study of the cluster include an
optical study by Jones \& Jones (1980), detection of the X-ray gas in
the cluster (Killeen \& Bicknell 1988), a catalog of dwarf galaxies
(Ferguson 1989, hereinafter F89), and a study of the dynamics of the
globular cluster system of the central elliptical galaxy, NGC 1399
(Grillmair 1992).

A relatively small number of these galaxies have measured redshifts,
however. A purpose of our current program is to begin to remedy this
problem and to commence a detailed study of the kinematics of the
galaxies in the Fornax cluster.

\bigskip
\bigskip
\centerline{2. OBSERVATIONS AND REDUCTIONS}
\bigskip
\centerline{\it 2.1 The Program Sample}
\bigskip

Since very few dE's in the Fornax cluster have published spectroscopic
data, we just set out  to observe all the brightest ($B_T < 16.5$) {\it
nucleated} dwarf ellipticals listed by  F89
in his study of the luminosity function of galaxies in Fornax.
In the following, program objects will be denoted by their number
in the F89 catalog.
The fact that our sample is biased in favour of bright,
nucleated dE's must be borne in mind when discussing their statistical
properties.
Only those galaxies positively identified as dE members of the
cluster ("class 1") were selected. Morphological membership criteria
proved to be highly successful.
In order to avoid duplicating existing observations, our
master list was cross-checked with:
(i) a copy of the Huchra's unpublished ZCAT redshift catalog;
(ii)  Fornax galaxies collected by  Brodie \& Huchra (1991).
These authors quote radial velocities and discuss metallicities of 10
galaxies in Fornax, 6 of which are dwarfs, from various unpublished
sources; (iii)  the sample of 5 dwarf galaxies observed and briefly
reported by Gregg (1992).

We observed with highest priority all the dE's in the CB87 list with
$B_{nuc} < 21$ (or correspondingly high central surface brightness).
Other high central surface brightness
candidates for spectroscopy were selected by short-exposure B imaging.
Table 1 contains the basic data of our spectroscopic sample of Fornax
dwarfs. Photometric data are from Caldwell (1987), CB87, and F89.
Although these galaxies are all classified as nucleated dE's, with the
only exception of FCC~261 (dE3pec,N/ImIV), they have rather different
surface brightnesses and structure (Table 1).

\bigskip
\bigskip
\centerline{\it 2.2 Spectroscopy}
\bigskip

Spectroscopic observations of nucleated dE's in the Fornax cluster have
been made with the ESO Multi Mode Instrument (EMMI) on the 3.5m NTT
telescope of the European Southern Observatory at La Silla, Chile.
The EMMI was operated in medium dispersion mode in the blue channel,
where the detector was a thinned TEK $1024\times1024$ CCD
(ESO No. 28). CCD read-out was restricted to a $1100 \times 601$ region
to minimize overhead time, implying a slit length of $3'.6$.
The scale along the direction of the slit was $0".36$ per pixel.
Long-slit spectra were obtained with a 300 grooves/mm grating blazed
at 4000 \AA\ (ESO No. 6), yielding a dispersion of 72 \AA\ mm$^{- 1}$ (or
1.73 \AA/pixel) over the spectral range $\sim 3700-5200$ \AA\ (the upper
limit was set by the EMMI's blue arm response).
With a $1".2$ slit, the effective resolution was $\sim 5.5$ \AA\ (FWHM)
($\sim 3$ pixels or $\sim400$ \kms at $\lambda$ 4000 \AA\ ), as measured
directly on the final reduced spectra.

Program galaxies were observed on two clear nights of 1-2 December
1991, with typical exposure times 1800--3600s in
single exposures (except for the galaxy FCC~207 whose observation was
split in two 1800s exposures).
This corresponds to $\sim$800 to $\sim$1500 photons per pixel at 4800 \AA\
in
most extracted spectra. FCC~85 and FCC~243 are of poorer quality
(300--400 photons pixel$^{\rm -1}$), which has led to rather uncertain
results for these two dwarfs.
We also obtained spectra of the Galactic globular clusters NGC~104 (47
Tuc) and NGC~2808, which provide useful reference [Fe/H] values for
checking our measurements of dE's metallicities. As well, these clusters
are good radial velocity templates (Webbink 1981).
The telescope was nodded in a direction perpendicular to the slit to
sample the average population of their centers.
Spectra of a few velocity and flux standard stars were also taken
in twilight.

Data analysis was carried out with a DEC/5000 workstation at the Bologna
Astronomical Observatory using the IRAF
\footnote{$^2$}
{IRAF is distributed by the National Optical Astronomy Observatories,
which is operated by the Association of Universities for Research in
Astronomy, Inc. (AURA) under cooperative agreement with the National
Science Foundation.}
package.
The data frames were bias-subtracted, and flat-fielded using quartz lamp
dome flats for correcting pixel-to-pixel sensitivity variation.
The spectra were then extracted using the variance-weighted extraction
algorithm of IRAF, which achieved an effective cleaning of
cosmic ray hits. Spectra were also visually inspected and a few residual
bad pixels interactively interpolated over.
Integrated spectra of 47~Tuc  and NGC~2808 were obtained by co-adding a
$\sim 25"-30"$ region across the core, with the sky taken at the edge of
the slit farthest from the globular cluster center.
Object+sky spectra were also extracted for the dE's
using the same parameters as for the object spectra, aside from
sky subtraction. They were used to estimate errors from photon
statistics.
He-Ar lamp exposures taken before and after each object spectrum were
used for wavelength calibration, with a solution error of 0.02--0.07
\AA\ (1.5--5 \kms, $1 \sigma$). The raw spectra were then rebinned to a
linear wavelength scale of 1.5 \AA\ per pixel.
For measurement of metallicity indices, we also produced a set of
spectra in relative flux units, using observations of a
spectrophotometric standard star (L745-46A), and mean extinction
coefficients.
Since absolute spectrophotometry is beyond the scope of this paper, no
attempt was made to put the spectra onto an absolute flux scale.

The wavelength-calibrated spectra of Fornax dE's and template clusters
are shown in Figure 1 on a relative linear flux scale.
They are arranged roughly in order of increasing strength of the 4000 \AA\
break (see Section 3.2), and smoothed with a 3-pixel kernel.
The absence of \CAII H, K in FCC~85 is attributed to the poor signal.

The IRAF cross-correlation task FXCOR was used to measure the shift in
velocity of the dE's with respect to the velocity standard
stars and globular clusters.
Internal velocity errors were computed following the precepts of
Tonry
\& Davis (1979). These do not include the contribution of systematic
errors, such as the uncertainty in the wavelength calibration of
program objects and templates.
To evaluate the accuracy of the wavelength scale, we have directly measured
template velocities by fitting a Gaussian profile to 6--7  absorption lines
in
the template spectra.
The uncertainty is of the order 20 \kms or less, $i.e.$ $\sim 1/20$ of the
resolution element (FWHM).
For each galaxy, the velocities obtained from 3 different templates are
consistent to within 1/3 of the internal error.
Heliocentric velocities are also given in Table 1; they confirm
the membership of all of our candidates.

\bigskip
\bigskip
\centerline{3. ANALYSIS OF THE SPECTRA}
\bigskip
\centerline{\it 3.1 Measurement of Spectral Indices}
\bigskip

Line strengths have been measured in Galactic and extragalactic
objects by several authors.
Brodie \& Huchra (1990; hereafter BH) have recently
collected and discussed a number of indices defined  by Burstein \etal
(1984) and Brodie \& Hanes (1986),
and derived empirical [Fe/H] calibrations based on the
Galactic globulars' metallicity scale of Zinn \& West (1984).
The index selection of BH is of special interest for us for two reasons.
First, their spectra have resolution very similar to ours, which allows
us to retain their bandpass definitions (repeated in Table 2
for convenience).
Indices in this first group comprise CH (G band) and H$\beta$ from
Burstein \etal (1984),
H$+$K and $\Delta$ (a color measuring the 4000 \AA\ break) from
Brodie \& Hanes (1986).
We note that unfortunately the red cutoff of
the blue EMMI spectra prevented measurements of the strong Mg2 feature.
Second, we can use Brodie \& Huchra's  [Fe/H] calibration of metal line
indices to estimate the  metallicity of Fornax dwarfs.
For this reason we have closely followed all BH reduction procedures.
The metallicities derived from our spectra of 47 Tuc and NGC~2808
have provided a useful (although limited) check {\it a posteriori} of the
adopted [Fe/H] scale.
%

We have also examined indices defined by other authors
(Stetson 1984; Rich 1988; Da Costa \& Mould 1988; Gregg 1989).
Feature and continuum band widths have been adapted to the spectral
resolution of our data.
Among line indices discussed in the literature, sidebands were
chosen to be (i) relatively free from absorption line
contamination, and (ii) close to each measured line, to make indices
as independent as possible of continuum slope.
Our new K, Fe {\sc i}~4384, H$\gamma$, H$\delta$, and H8 (H$\zeta$)
indices are defined in Table 2. In particular, H8 was defined so as to
be little affected by the adjacent CN band with head at 3883 \AA.
The $\Delta^\prime$ index is similar to $\Delta$ but excludes the
noisy region blueward of 3850 \AA\ .

Following BH, line indices were defined as

$$ I = - 2.5 * log_{10} \left( F_l \over F_c   \right) , \eqno(1) $$

where $F_l$ and $F_c$ are the mean fluxes in the line and continuum
bandpasses. Pseudo-equivalent widths can be easily calculated
from magnitude indices  by the relation:

$$ W_\lambda = \Delta\lambda * \left( 1 - 10^{-0.4*I}  \right) , \eqno(2)$$

where $\Delta\lambda$ is the feature bandwidth.
The indices were measured by simply integrating the flux within the line and
continuum bandpasses in the smoothed, flux--calibrated spectra.
(The 3-pixel boxcar smoothing we applied to our spectra has little
effect on the indices. In fact, line measurements on the fluxed,
unsmoothed spectra yielded the same results at the 1\% level).

The bandpasses were suitably redshifted to account for the measured
velocities.
Where two sidebands are defined, we estimated the continuum at the
line center by linearly interpolating the average continuum levels in
the sidebands. For the indices with only one continuum bandpass, the
average flux in that bandpass was taken.

Extensive tests have shown that
these indices are essentially independent of the reduction
procedure, or of the intrinsic continuum slope.
Only $\Delta$, \CAII K and H8 are somewhat dependent upon
the local continuum.
The values of the indices are listed in Table 3.
 Errors in Table 3
 were estimated (from read-out noise and photon statistics only) by
 measuring the auxiliary object+sky spectra in the same bandpasses.
%
Note that indices for the Galactic globulars NGC~2808 and 47~Tuc
are probably more uncertain than indicated by their formal errors, which
do not take into account inadequate sky subtraction.

Correction of line indices for the effects of internal velocity
dispersion ($\sigma_v$), that may be significant for giant
ellipticals, should be negligible here, since
$\sigma_v$ is expected not to exceed $\sim 50$ \kms
($e.g.$, Carter \& Sadler 1990; Held \etal 1990,1992; Bender \etal 1991;
Peterson \& Caldwell 1993).

\bigskip
\bigskip
\centerline{\it 3.2 Metallicities}
\bigskip
Given the fairly modest S/N ratio of our spectra of dE's,
several indices in Table 3 were combined to produce good estimates of
metallicity and hydrogen line strength.  The signal-to-noise ratio of
our spectra is certainly not sufficient to address the question of
relative abundances of light and Fe-peak elements, shown to be
variable in giant ellipticals (Peletier 1989; Worthey \etal 1992).
An Fe deficit (or Mg excess) is not present, however, in weak-line
low-luminosity-ellipticals (LLE), showing solar abundance ratios.

Figure 2 shows that the metal-line indices H+K, K, and Fe~I
correlate quite well with each other and the color $\Delta$,
as in Galactic globular clusters.
Starred symbols represent the two Galactic globulars.
The poor-S/N dE's FCC~85,
FCC~243, and (to a lesser degree) FCC~261 yield the most uncertain
indices. The correlation between $\Delta^\prime$ and $\Delta$,
not displayed here, is as good as expected.
Errors bars representative of statistical uncertainties in Table 3, are
shown for the best and the worst data point.
In most cases,
we attribute the outliers in these diagrams to the unaccounted errors
related to poor sky subtraction, that are certainly most severe,
although not easily quantified, for the faintest spectra.
We note that the G-band strength is relatively insensitive to
metallicity in this range of magnitudes
(Burstein \etal 1984, Gorgas \etal 1993).
The CH index will not be further discussed here.

The correlations shown in Figure 2 were used to combine all the
metal-line indices into an average metallicity index.
Straight lines were fitted to the indices in Figure 2, yielding the
following relations:
$$ \Delta = 0.146 + 1.162 * HK $$
$$ HK = 0.058 + 0.507*K $$
$$ \Delta = 0.294 + 2.218 * FeI ,$$

also shown as straight lines in Figure 2.
A program was used that fits a straight line to data with errors
in both variables (based on the Lybanon's 1984 algorithm, implemented
and kindly made available to us by R. Vio). A few outliers were completely
excluded from the fit: 4 for H+K $vs.$ $\Delta$; 2 for K $vs.$ H+K; 1
(NGC~2808) for Fe~I $vs.$ $\Delta$.
This is based on the assumption that indeed
metallicity is the factor that drives these correlations.
In order to avoid giving too much weight to the globulars on the basis
of their (negligible but probably underestimated) formal errors, they
were assigned weights equal to those of the best measured dE's.
Following the approach of Brodie \& Hanes (1986), H+K, K, Fe~I, and
$\Delta^\prime$ were transformed into equivalent $\Delta$ values.
Average $\Delta$ indices were then obtained, weighted by the inverse
square of the statistical errors; the outliers were discarded, which is
almost equivalent to taking the median of the $\Delta$'s.
The reason for reducing all indices to an equivalent value of $\Delta$
is that, among the indices in Table 3, $\Delta$ is the only one for
which BH provide a metallicity calibration independent of the object
type. For H+K, different calibrations have been found for Galactic and
M31 globulars, probably due to abundance anomalies (Brodie \& Huchra
1991).

The metallicities obtained by applying the Brodie \& Huchra's (1990)
$\Delta$--[Fe/H]
calibration (for "All" objects: [Fe/H] = 3.18 * $\Delta$ - 2.43) to our
weighted $\Delta$'s are listed in column 3 of Table 4. In column 2 is
the number of indices averaged for each object (less than 5 if any
discrepant indices were dropped, usually H+K or K of \CAII).
Errors were estimated from the (weighted) standard deviation of
$\Delta$'s from the average value. They were not divided by the square
root of n$_{\rm [Fe/H]}$. It should be borne in mind that these errors
merely reflect the internal uncertainties affecting the mean $\Delta$
index, and do not account for uncertainties in absolute
calibration.
Also given in Table 4 are [Fe/H] values obtained from the median of all
(five) $\Delta$ values. As expected, the two abundance determinations
are in excellent agreement, with the exception of NGC~2808.
The errors on the median are simply the standard (unweighted) deviation
of all data from the median itself. Therefore they are likely to place an
upper limit to the internal uncertainties of our measurements.
The errors quoted in column 3 appear more optimistic as
a result of weighting.

The metallicity ranking of Table 4 was also checked by eye on the
spectra in Figure 1.
The metallicities in Table 4 are confirmed, except for FCC~243 that
looks much more similar to metal-poor dwarfs ($e.g.$, FCC~188, [Fe/H]$\sim
-1.25$) than to metal-rich ones. Its high measured metallicity, reported in
Table 4, is likely to be an artifact of the deepness of \CAII H and K
lines (Figure 1), for which we could not find any obvious explanation.
In the following, we have adopted for FCC~243 our visual estimate, with
the uncertainty quoted in Table 4.

How reliable is our adopted metallicity scale ?  Since we have closely
followed Brodie \& Huchra (1990) in measuring the indices, a systematic
offset seems unlikely, though it cannot be ruled out {\it a priori}.
Indeed, the [Fe/H] values obtained for both NGC~2808 and 47~Tuc
turn out to be very close to the published values (-1.37 and -0.71,
respectively -- Zinn \& West 1984).
Further, the mean $\Delta$ indices of the two template globular clusters
essentially bracket the range of values for Fornax dE's, so that we
were able to assign metallicities to the dE's without extrapolation.
Thus, while a larger control sample would certainly be desirable,  we
believe that the metallicity scale presented in Table 4 is
basically consistent with Brodie \& Huchra's, within say 0.1--0.2 dex in
[Fe/H].

This is confirmed by the agreement between metallicities obtained
in this paper and Brodie \& Huchra's (1991) results for Virgo and Fornax
dwarf ellipticals.
These authors found that (normal and dwarf) elliptical galaxies conform
well to the index-metallicity relations defined by globular clusters,
and therefore applied their GC metallicity calibration also to galaxies.
Their empirical procedure implicitly assumes that stellar populations of
galaxies are (at least roughly) similar to those of Galactic and M31
globulars. This is not generally true for dwarf ellipticals, some
of which are known to harbour a young or intermediate-age stellar
population.
The features in the UV-blue spectral region are expected to
be sensitive not only to metallicity, but also to the age of the last
star formation episode. This is an aspect of the well-known difficulty
in distinguishing between the effects of age and metallicity
from data in a limited frequency range, for example blue spectra (\eg
Rabin 1982).
Recent studies of normal E/S0's provide circumstantial evidence that
some features
are indeed sensitive to {\it both} metallicity {\it and} age.
%
%
%
For instance, the 4000 \AA\ break, as well as the
\CAII H, K and Fe~{\sc i} 4384 \AA\  lines, which contribute to
defining the metallicity ranking of dE's in this paper, are weak at a
given Mg strength in star-forming ellipticals and some S0 disks
(Kimble \etal 1989; Gregg 1989).
%
This suggests the possibility that
metallicity measurements of some dE's from our sample -- essentially
those with strong Balmer lines -- might be subject to similar deviations
($cf.$ Gregg 1992).
With these cautionary remarks in mind, we use the Brodie \&
Huchra's (1990) [Fe/H] scale as a good approximation to metallicities of
dE's, under the assumption that our indices depend primarily (though not
solely) on metal abundance. We shall return to this point in our subsequent
discussion.

\bigskip
\bigskip
\centerline{\it 3.3 Hydrogen Line Strengths}
\bigskip

A similar procedure was applied to combine low-S/N Balmer line indices
into a mean hydrogen line strength
\hmed $ =
{1 \over 3} ({\rm H}\beta+{\rm H}\gamma+{\rm H}\delta)$,
an index also employed in studies of high redshift galaxies.
The H$\beta$, H$\gamma$, and H$\delta$ indices were plotted against each
other, and regressions obtained using the same techniques as for
metallicity.
For some dE's, the H$\beta$, H$\gamma$, and H$\delta$ line strengths are
well correlated, so that the three lines were easily averaged. From
these data, we also derived scaling relations between
the strength of each line and \hmed.

However, some dE's have discrepant values of either H$\gamma$ (FCC~296)
or H$\delta$ (FCC~85, FCC~207), which are generally attributed to the
low S/N of the spectra.
Further, FCC~243 and FCC~261 have H$\beta$ too weak for their
H$\gamma$ and H$\delta$ (the pseudo-equivalent width
of H$\beta$ alone is $0.35 \pm 0.85$ and $0.43 \pm 0.57$, respectively).
This may indicate the effects of incipient Balmer emission lines, too
weak to be detectable as a reversal, yet causing some filling of the
$H\beta$ absorption lines.
In all these cases, weighted average values were obtained of the two
consistent lines, and the result reduced to \hmed
using the scaling relations described above.

The average H line strengths are listed in column 6 of Table 4.
Note that our choice of bandpasses for Balmer lines gives slightly lower (by
20\%) pseudo-equivalent widths than other definitions in the literature
(\eg Caldwell \etal 1993). The range of mean H line strengths of dE's,
1-3 \AA, found here is also typical for spiral galaxies with similar
colors.
Errors were calculated from both count statistics and the average
(weighted) variance of Balmer indices. Only for the globulars 47~Tuc and
NGC~2808 is the measured scatter of the indices larger than
would be expected from photon statistics. The larger of the two error
estimates is given in Table 4.

Caldwell \etal (1993) have recently pointed out that W(H8) is a more
reliable indicator of the presence of hot stars than H$\delta$, since
it is less sensitive to filling by Balmer emission (we are grateful
to the referee for calling our attention on this diagnostic).
The pseudo-equivalent widths of H8 are listed in column 7 of Table 4
along with their statistical errors.

\bigskip
\bigskip
\centerline{4. DISCUSSION}
\bigskip
\centerline{\it 4.1 Stellar Populations}
\bigskip

The present
analysis of the spectra of dE's has the advantage of providing a
two-dimensional classification not available from integrated colors,
although still semi-qualitative owing to the lack of proper spectral
synthesis models with a full grid of ages and metallicities.
In a coeval population, hydrogen lines sample the warm stellar component,
particularly the temperature distribution on the main-sequence turn-off and
the subgiants ($e.g.$, Buzzoni \etal 1993).
Figure 3 shows the Balmer line equivalent widths of dwarf ellipticals
and Galactic globulars in a (metals, hydrogen) diagram similar to the
age-metallicity diagnostic diagram of Rabin (1982). FCC~243 has been
plotted using its visually estimated metallicity. The error bars
represent the uncertainties quoted in column 4 and 6 of Table 4.
In this figure, the average \hmed  strengths have been scaled (by a
factor 1.2) to the pseudo-equivalent widths of H$\beta$ alone, to
facilitate the comparison with globular cluster data.
The reference line in Figure 3 is the locus of Galactic globular
clusters, interpreted as  a sequence of varying metallicity at constant
age. This is defined by the least-squares fit to H$\beta$ strengths
versus Galactic globular cluster metallicities presented in Brodie \& Huchra
(1990). No rescaling or adjustment has been applied.
It should be reminded here that the results described in this paper
refer to the nuclei of dE's. Spectroscopy of non-nucleated dE's, or of the
external regions of nucleated dwarfs, is made exceedingly difficult by
their extremely low surface brightness.

%
A number of dwarfs in our sample (FCC~85, FCC~100, FCC~188, and
FCC~296) have mean hydrogen line strengths consistent with or marginally
weaker than those expected for an old, metal-poor population.
In contrast, there are some dE's showing  enhanced hydrogen lines relative to
globular clusters of similar metallicity. All the hydrogen lines, except
for H$\beta$, are particularly enhanced in FCC~261.
The Balmer lines are also moderately strong, \hmed $> 1.5$ \AA, in FCC~243,
FCC~245, FCC~266, and FCC~150 (in order of decreasing intensity).
For all of these dwarfs, the equivalent width of H8 is also larger than 3 \AA.
H$\delta$ is abnormally strong ($\sim 3$ \AA) relative to the other H
lines in the spectrum of FCC~207.
For FCC~207 and FCC~296, the \hmed values in Table 4
and Figure 3 are likely to be underestimates of the H line strength,
since H8 indicates some hot star content.
For FCC~150 and FCC~266, which are relatively metal-rich, the hydrogen line
strengths are explained by a relatively blue stellar component. These
strong Balmer lines cannot be due to low metallicity. As discussed
below, the metallicities of these dwarfs may be, if any,
underestimated. The moderate intensity of the hydrogen lines in these
two dwarfs suggest the presence of an intermediate-age population,
rather than a recent burst of star formation ($cf.$, Bothun \& Mould
1988).
In the case of FCC~261 and (to a lesser degree) FCC~243 and FCC~245, the
hydrogen lines are strong enough to be consistent with the presence of a
young stellar component, even in the absence of distorted morphology.
In addition, recent star formation in FCC~261 is unambiguously confirmed
by a peculiar knotty appearance on blue images.

Hydrogen line strengths are more sensitive to a warm-star
component than are integrated colors. For instance, NGC~185 has not
abnormally blue $U - V$ color for its luminosity, even with its blue
stars. Also, dwarf ellipticals with morphological signatures of young stars
are only 0.1--0.2 mag bluer in $U - V$ than predicted from a
color-luminosity relation (Caldwell 1983).
In Figure 4
W$_{{\rm H}\beta}$ is plotted against $U - B$ for Fornax dwarfs
(only 6 dwarfs had their colors measured by CB87). Galactic globular
clusters are also plotted for comparison (colors are from Reed \etal
1988, H$\beta$ measurements from Brodie \& Huchra 1990). Figure 4
provides further evidence that Balmer lines are stronger in
FCC~245 than expected for an old, metal-poor population, and indicates
the need for a younger component with respect to globular clusters.

Thus, there is some evidence that young stars or an intermediate-age
main-sequence turn-off are relatively common among dE's, in accord with
a suggestion put forth by some authors (Bothun \etal 1985, Zinnecker
\etal 1985) that nuclei are the site of recent star formation.
The absence of peculiar morphology seems to suggest that residual star
formation must have concentrated in the nucleus.
An intermediate-age AGB population would also be expected.

Nearby Local Group dwarfs may provide a closer picture and helpful
guidelines on understanding the stellar population of dE's.
NGC~205 is known to be experiencing a modest burst of star
formation (Wilcots \etal 1990), and asymptotic giant branch stars are
observed in the central regions, which implies a significant
intermediate-age (0.5--1.5 Gyr) stellar population (Mould \etal 1984;
Richer \etal 1984; Davidge 1992).
An analogy with Local Group dwarf ellipticals would suggest that an
interpretation of Balmer line strengths in terms of coeval stellar
populations might not be appropriate. Instead, nucleated dE's may be
forming stars in bursts from the gas recycled from the old population,
as proposed by Gallagher \& Hunter (1981) for NGC~185 and NGC~205.

%
\bigskip
\centerline{\it 4.2 Metallicity}
\bigskip

As to metallicity, Figure 3 shows that Fornax dE's in this study are
distributed in a relatively wide range in [Fe/H], from moderately
metal-poor ([Fe/H] $\sim -1.4$ dex) to values similar to those of
metal-rich globular clusters (--0.7 dex).
In our limited luminosity range (M$_{\rm B} \sim -15 \pm 0.5 $ mag,
adopting a distance modulus 31.0 from Aaronson \etal 1980), we find an
apparent scatter of 0.23 dex around a median metallicity [Fe/H] =
$-1.20$, with no apparent correlation between [Fe/H] and luminosity.

[Fe/H] values derived from blue spectra are well correlated with $U - B$
colors from CB87. In Figure 5$a$, metallicities  are
plotted against $U - B$ colors for the Fornax dE's in common with CB87
and the globulars 47~Tuc and NGC~2808.
While this relationship is not unexpected, since U--B is sensitive to
metallicity through the strength of the 4000 \AA\ blanketing
discontinuity, this correlation has important implications.
First, it shows that our formal error bars are realistic, and
gives confidence in our metallicity measurements. More importantly, this
correlation demonstrates that the apparent range in metallicity at
a given luminosity is real, i.e. not caused by observational scatter.
This conclusion also applies to the range in $U - V$ colors discussed by CB87.
Consequently, the general metallicity-luminosity correlation can be
studied only statistically,
in a wide luminosity range.
Metallicity from blue spectra also correlate well with $B - V$ (Figure 5$b$).
There is no shift relative to globular clusters. This correlation
is an analog of that between $\Delta$ and a spectral gradient in normal
elliptical (Kimble \etal 1989), and is driven by the temperature of the
main-sequence turn-off.
We note that a correlation is also implied between $U - B$ and $B - V$.

FCC~207 is too blue in $U - B$ (by $\sim 0.1$ mag),
and apparently too metal-poor (by 0.15 dex) for its $B - V$.
These effects point towards a young stellar component.
Note that $U - B$ colors and [Fe/H] estimates are affected in a similar way,
so that FCC~207 lies on the GC relation in Figure 5$a$.
The only other dwarf with strong H lines $and$ published colors, FCC~245,
appears quite normal in Figure 5.

In terms of absolute [Fe/H] values, our results are tied to
the Brodie \& Huchra's metallicity scale.
We note the agreement with the metallicities of the dE companions to M31,
NGC~205 and NGC~147, directly obtained from color-magnitude diagrams
(Mould \etal 1984).
To use a calibration-independent statement, Fornax dE's have spectra
resembling those of intermediate metallicity to metal-rich globular
clusters.

This well established range in spectral characteristic is in good
agreement with previous results from blue spectra.
Spectroscopy of 3 dE's by Bothun \etal (1985) suggested
that a strong-lined Virgo dwarf, NGC 4472-DW8,
is more metal-rich than 47~Tuc (for this object Brodie
\& Huchra obtained [Fe/H] $= -1.2 \pm 0.5$, the two results being only
marginally consistent). Two other  Virgo dwarfs are either moderately
metal-poor (Fe/H $= -1.0$ to $-1.5$, NGC 4472-DW10) or very metal-poor,
based on their \CAII line strengths.
Bothun \etal (1985) noted that {\it the dE's
form a heterogeneous sample in terms of their spectroscopic properties}.
In their study of 12 nucleated Virgo dE's, Bothun \& Mould (1988)
concluded that blue nuclear spectra resemble those of
metal-rich globular clusters, though some dE's are probably more
metal-poor (yet considerably more metal-rich than M92). Indeed, we note
that most of the Virgo dwarfs in the Bothun \& Mould's sample have \CAII
K line strengths between those of M79 ([Fe/H] $= -1.68$) and 47~Tuc.
Most recently, Brodie \& Huchra (1991) obtained, from 5 blue-visual
spectral indices, a mean metallicity of --1.15 for Virgo and Fornax dE's.
They pointed out a discrepancy between metallicities derived from
optical line strengths and those derived from infrared colors.  For
example, the $J - K$ colors, which are sensitive to metallicity through
giant branch temperatures, indicate metallicities $\sim
-0.7$ for Virgo dE's (Bothun \etal 1985; Thuan 1985).

It would have been of interest to compare our new optical results to
IR data. Unfortunately, no infrared photometry has been published
in our knowledge for Fornax dE's.
A direct comparison of $J - K$ colors and \CAII K line strengths is
however
possible for a small sample of Virgo dE's with IR data (Bothun \etal 1985,
Thuan 1985) and spectroscopic data (Bothun \& Mould 1988)
(Figure 6).
These are
M87-- DW1, DW6, DW11, DW27, and DW31.
Note the $\sim 0.1$ mag offset in the $J - K$ colors of Thuan (1985)
relative to those of Bothun \etal (1985) and Zinnecker \etal (1985).
Figure 6 suggests that $J - K$ colors are independent of the strength of
the \CAII K line.  For dE's with strong \CAII K, infrared colors and
spectra are consistent, yielding abundances similar to those of
metal-rich globular clusters. For weaker-lined dwarfs, $J - K$ and line
strengths in the blue spectra apparently give divergent results, in that
$J - K$ colors are redder than expected for an old, moderately
metal-poor stellar population.
Given the correlation in Figure 5$a$, this is equivalent to the
photometric result that dE's have bluer $U-B$ colors than globular
clusters of the same $J -K$ (CB87).
Clearly, it would be important to perform this exercise on a larger sample
of dwarfs (including those in this paper) having both accurate IR colors
and homogeneous spectroscopy.

To explain this discrepancy, Brodie \& Huchra (1991) argued
that the IR color-metallicity calibration based on Galactic globular
clusters may not be appropriate for dwarf ellipticals. Differences in
stellar population mix may be present in dE's compared to the globulars. In
particular, $J - K$ colors could be made redder by a population of AGB
stars. Bothun \etal 1985 pointed out a problem with
the $H - K$ colors, apparently consistent with those of globular
clusters. However, plotting $H - K$ colors against \CAII K for the same
sources as in Figure 6, we have found that dE's have slightly {\it
redder} $H - K$ than globulars of similar \CAII line strength. Thus we
believe AGB stars are a possibility not to be ruled out, but clearly
more data are required.

We now examine the alternative possibility that metallicity
as derived from blue spectroscopy is biased.
CB87 suggested that the range in optical colors of Fornax dE's
could be partly explained by a dispersion in the mean ages.
Similarly,
perhaps the scatter in [Fe/H] measured from blue spectra is not
entirely due to metallicity.
As shown in the previous section, it is conceivable that metal line
strengths of some dwarfs are diluted by a UV light excess, which
potentially leads to an underestimate of metallicity.
Indeed, simulations of a recent burst in giant ellipticals by Bica \etal
(1990) show that \CAII K is quite sensitive to dilution effects by the
UV continuum, even for bursts involving only 0.1 \% of the galaxy mass -- a
result expected to hold at least qualitatively also for metal poorer dwarf
ellipticals.

The dwarfs with strong hydrogen lines are the first candidates for such
effect, since both Balmer lines and $\Delta$ (as well as blue-visual
colors) are controlled by the temperature of the main-sequence turn-off.
For example, the metallicities of FCC~261 and FCC~245 might be too low
by a few tenths of a dex,
and "corrected" data points should be moved to the right
(higher metallicities) in Figure 3. As an important consequence, Balmer
line strengths would be even more in excess with respect to globular
clusters. Evidence for a young or intermediate-age population would be
strengthened.

However, the observed metallicity range is unlikely to be
entirely explained by dilution effects, for the following reasons:
(i)
The same range of metallicities as found here is given by Brodie \&
Huchra (1991) using also indices at longer wavelengths, such as Mg2 or
Fe52. The dependence of Mg2 on metallicity is well studied (Mould 1978;
Worthey \etal 1992; Buzzoni \etal 1992; and references therein). As
regards Fe52, this index mainly reports on the metallicity of the
moderately cool stars at the base of the RGB, with little dependence on
age (Buzzoni \etal 1993).
(ii)
Since blue indices are likely to be modified at a different degree,
dilution effects should be apparent in index-index plots. In fact,
Figure 2 shows that FCC~261, the only case where a blue light excess is
evident, has H+K and K indices consistently too strong for its $\Delta$.
Further, the G band is too weak (CH is otherwise nearly constant in this
abundance range -- Burstein \etal 1984). Differential dilution effects
are, however, not evident for other objects with strong Balmer lines.


In Figure 7, we have also looked at a possible correlation between
metallicity and surface brightness for dE's (the effect of a threshold
gas density for star formation has been discussed, $e.g.$, by Bothun \&
Mould 1988 and Phillipps \etal 1990).
This figure shows no clear correlation between metallicity and effective
surface brightness; a similar conclusion is drawn from a plot of
[Fe/H] against nuclear magnitude, or the average central surface
brightness S3pix (both from CB87).
However, we note that the two dwarfs with the highest surface
brightnesses, FCC~150 and FCC~266, also have high metallicities
($cf.$ NGC 4472-DW8 in Bothun \etal 1985).

\bigskip
\bigskip
\centerline{\it 4.3 Kinematics of the Fornax Cluster}
\bigskip

The distribution of the dwarf galaxies in Fornax is shown in Figure 8.
In addition to the ten galaxies in Table 1 there are 85 galaxies
in ZCAT (Huchra 1990) which satisfy the following criteria:\hfil\par
$\bullet$ Redshift $cz~<$ 2520 \kms \hfil\par
$\bullet$ blue magnitude
\footnote{$^3$}{We do not distinguish for present purposes between
B$_T$ and m$_B$}
known (de Vaucouleurs \etal 1990, F89)\hfil\par
$\bullet$ Location within 5$^\circ$ of NGC 1399.

The mean heliocentric velocity of this sample is 1450 \kms and the
velocity dispersion is 330 \kms. If the core radius ($r_c$) of the
cluster
is 0.7$^\circ$ (F89) and the distance modulus 31.0 (Aaronson  \etal 1980),
the crossing time ($r_c/\sigma$) is 0.5 Gyr.
Figure 9 shows no evidence for organized motions
in the velocity distribution of the sample galaxies, and, in particular,
there is no evidence for rotation of the cluster about the minor axis of
the
distribution in Figure 8 at the level $v/\sigma~<$ 0.07.

There is a detectable difference, however, between the velocity dispersion
of the dwarf galaxies (B$_T~>$ 15 mag) and the rest of the sample. This
is indicated in Table 5. Figures 10 (a) and (b)
illustrate the normal distribution
of the brighter sample and the flatter distribution of the dwarf sample.

The Kolmogorov-Smirnov test rejects the hypothesis that the dwarf sample
is drawn from a Gaussian distribution with a dispersion of 300 \kms
at the 75\% confidence level. This is not sufficient for us to be able
to conclude that there is a difference in the kinematics of the two
samples,
but it is indicative that there is some probability that this is the case.
A similar flatter distribution was seen in a Virgo sample of dwarf galaxies
by Bothun \& Mould (1988).
A larger Fornax sample should be investigated in order to confirm these
interesting, but preliminary, results.

It is of interest, nonetheless, to ask how such velocity dispersion
differences might physically arise. To discuss this, we recall
the relevant timescales. Although cluster virialization may
be established in a few crossing-times, and the latter, as indicated above,
is short, cluster growth occurs on the infall timescale. The infall
timescale for mass shells to turn around and accrete into a cluster
is of order 5 Gyrs for the Coma cluster (Gunn \& Gott 1972). This
collapse time scales like the crossing-time, which is similar
for the two clusters.
The timescale for mass segregation, however, is the relaxation time,
which, even in a cluster as dense as Fornax, exceeds the Hubble time
by an order of magnitude.

The candidate process, therefore, for velocity dispersion differences
in clusters of galaxies, is accretion of galaxies over the lifetime
of the cluster. The morphology-density relation (Dressler 1980) would
tend produce an initial distribution of galaxies in which dwarf galaxies
preferentially occupy the outer halo of the proto-Fornax cluster,
while massive early-type galaxies occupy the core. The longer infall
timescale of the halo dwarfs would then yield a larger velocity dispersion
characteristic of their more recent separation from the Hubble flow.

\bigskip
\bigskip
\centerline{5. SUMMARY}
\bigskip

Spectra of the nuclei of ten dE's in the Fornax cluster have been
obtained, from which radial velocities have been measured and spectral
indices derived, related to either metal or hydrogen line strength.
Using the empirical Brodie \& Huchra's (1990) calibration based on
Galactic and M31 globular clusters, metallicities have been derived for
the dwarf ellipticals in our sample.
In this paper we have found a large variety of spectral characteristics
among the nuclei of dwarf ellipticals, although all targets (except
FCC~261) were selected on the basis of a uniform dE,N morphological
classification.

1. Blue spectra of dE's cover a wide range of metallicities,
from {\it metal-rich} ([Fe/H]$\sim -0.7$) to {\it intermediate metallicity}
globular clusters ([Fe/H]$\sim -1.4$), with some concentration
around [Fe/H] $\sim -1.2$ and $-0.7$.
This is in agreement with the metallicities
derived using color-magnitude diagrams for the dE companions to M31
(NGC~205 and NGC~147) and with previous spectroscopic results for Virgo and
Fornax dwarfs. However, it also confirms a discrepancy with metal
abundances inferred from infrared colors, at least for the
weaker-lined dwarfs. Possible reasons for this discrepancy have been
discussed, including contribution of an intermediate-age population to the
IR colors, and dilution of spectral features by a blue light  excess.

2.
A wide range in hydrogen lines strengths is highly suggestive of a
complex star formation history in dwarf ellipticals. In particular, a
hot stellar population is shown in some (but not all) nucleated dE's.
Some metal-rich dE's have on average relatively strong H lines,
consistent with an intermediate age population, but not as intense as
those found in NGC~205, where recent star formation is evident.
Other dwarfs (\eg FCC~245, FCC~261) show  stronger H lines,
consistent with recent star formation. For FCC~261, this is confirmed by
a small-scale structure on blue images.
A few relatively metal-poor dE's have hydrogen lines consistent or
perhaps weaker than those of Galactic globulars of comparable metal line
strength.

3. There is no clear correlation between metallicity and surface brightness
for the dwarf ellipticals having $\mu_B > 23.4$ mag arcsec$^{-2}$. However,
FCC~150 and FCC~266,  the two dE's with the highest surface brightness in
our sample, are both similar to metal-rich globular clusters. These objects
may represent intermediate cases between dE's and low-luminosity "normal"
ellipticals.

4. The radial velocity distribution of Fornax dwarfs looks flatter than
that of brighter galaxies. Numbers are small, however.
The hypothesis that both samples are drawn from the same parent
population can be rejected at the 75\% confidence level, which is
indicative of (yet does not prove) a difference in the kinematics of the
two samples.

These suggestions are important for their implications for stellar
populations in dwarf ellipticals. The trends are weak, however, due to
the small number of objects, and more data are needed.

\vskip 1.truecm
We are grateful to the Italian C.N.R.
for supporting one of us (JRM) during a visit
to the Osservatorio Astronomico di Bologna. We are obliged to John
Huchra for a recent copy of ZCAT. We would also like to thank the ESO
support staff for their assistance.
We are indebted to an anonymous referee for many useful suggestions.
This research has made use of the NASA/IPAC Extragalactic Database (NED)
which is operated by the Jet Propulsion Laboratory, Caltech, under contract
with the National Aeronautics and Space Administration.

\vfill\eject
\parindent=0pt

\bigskip
\bigskip
\centerline{REFERENCES}
\bigskip

\ref
Aaronson, M., Dawe, J., Dickens, R., Mould, J., \& Murray, J. 1980,
MNRAS, 195, 1P

\ref
Bender, R., Paquet, A., \& Nieto, J.-L. 1991, A\&A, 246, 349

\ref
Bica, E., Alloin, D., \& Schmidt, A. 1990, MNRAS, 242, 241


\ref
Bothun, G. D., \& Caldwell, C. N.  1984, ApJ, 280, 528

\ref
Bothun, G. D., Mould, J. R., Wirth, A., \& Caldwell, N. 1985, AJ, 90, 697


\ref
Bothun, G. D., \& Mould, J. R. 1988, ApJ, 324, 123

\ref
Brodie, J. P., \& Hanes, D. A. 1986, ApJ, 300, 258

\ref
Brodie, J. P., \& Huchra, J. P. 1990, ApJ, 362, 503

\ref
Brodie, J. P., \& Huchra, J. P. 1991, ApJ, 379, 157


\ref
Burstein, D., Faber, S. M., Gaskell, C. M., \& Krumm 1984, ApJ, 287, 586

\ref
Buzzoni, A., Gariboldi, G., \& Mantegazza, L. 1992, AJ, 103, 1814

\ref
Buzzoni, A., Mantegazza, L., \& Gariboldi, G. 1993, AJ, in press

\ref
Caldwell, N. 1983, AJ, 88, 804

\ref
Caldwell, N. 1987, AJ, 94, 1116

\ref
Caldwell, N.,  \& Bothun, G.D. 1987, AJ, 94, 1126 (CB87)

\ref
Caldwell, N.,  Rose, J. A., Sharples, R. M., Ellis, R. S., \& Bower R.
G. 1993, AJ, 106, 473

\ref
Carter, D., \& Sadler, E. M. 1990, MNRAS, 245, 12P

\ref
Da Costa, G. S. 1992, in IAU Symp. 149, The Stellar Population of Galaxies,
edited by B. Barbuy \& A. Renzini (Kluwer, Dordrecht), p. 191

\ref
Da Costa, G. S., \& Mould, J. R. 1988, ApJ, 334, 159

\ref
Davidge, T. J. 1992, ApJ, 397, 457

\ref
de Vaucouleurs, G., de Vaucouleurs, A., Corwin, H., Buta, R.,
Paturel, G., \& Fouqu\'e, P. 1992, Third Reference Catalogue of
Bright Galaxies (Springer-Verlag, New York)

\ref
Dressler, A. 1980, ApJ, 236, 361


\ref
Ferguson, H. C. 1989, AJ, 98, 367 (F89)

\ref
Ferguson, H. \& Sandage, A. 1988, AJ, 96, 1520

\ref
Gallagher, J. S., \& Hunter, D. A. 1981, AJ, 86, 1312


\ref
Gorgas, J., Faber, S.M., Burstein, D., Gonzalez, J., Courteau, S., \&
Prosser, C. 1993, ApJS, 86, 153

\ref
Gregg, M. D. 1989, ApJS, 69, 217

\ref
Gregg, M. D. 1992, in IAU Symp. 149, The Stellar Population of Galaxies,
edited by B. Barbuy \& A. Renzini (Kluwer, Dordrecht), p. 426

\ref
Grillmair, C. 1992, Ph.D. thesis, Australian National University

\ref
Gunn, J. \& Gott, R. 1972, ApJ, 176, 1

\ref
Held, E. V., Mould, J. R., de Zeeuw, P. T. 1990, AJ, 100, 415

\ref
Held, E. V., J. R., de Zeeuw, T., Mould, J. R., \& Picard, A. 1992, AJ,
103, 851

\ref
Huchra, J. 1990, priv. comm.

\ref
Jones, J. \& Jones, B. 1980, MNRAS, 191, 685

\ref
Killeen, N. \& Bicknell, G. 1988, ApJ, 325, 165

\ref
Kimble, R. A., Davidsen, A. F., \& Sandage, A. R. 1989, Ap\&SS, 157, 237


\ref
Lybanon, M. 1984, Am. J. Phys. 52, 22

\ref
Mould, J. 1978, ApJ, 220, 434

\ref
Mould, J., Kristian, J., \& Da Costa, G. S. 1984, ApJ, 278, 575

\ref
Peletier R. 1989, Ph.D. thesis, Rijksuniversiteit Groningen

\ref
Peterson R. C., \& Caldwell, N. 1993, AJ, 105, 1411

\ref
Phillipps, S., Edmunds, M. G., \& Davies, J. I. 1990, MNRAS, 244, 168

\ref
Rabin, D. 1982, ApJ, 261, 85

\ref
Reed, B. C., Hesser, J. E., Shawl, S. J. 1988, PASP, 100, 545


\ref
Rich, R. M. 1988, AJ, 95, 828

\ref
Richer, H. B., Crabtree, D. R., \& Pritchet, C. J. 1984, ApJ, 287, 138



\ref
Stetson, P. B. 1984, PASP, 96, 128

\ref
Thuan, T. X. 1985, ApJ, 299, 881

\ref
Tonry \& Davis 1979, AJ, 84, 1511


\ref
Tully, R. B. 1988, Nearby Galaxy Catalog (Cambridge Univ. Press., Cambridge)

\ref
Webbink, R. F. 1981, ApJS, 45, 259

\ref
Wilcots, E. M., Hodge, P., Eskridge, P. B., Bertola, F., \& Buson, L.
1990, ApJ, 364, 87

\ref
Worthey, G., Faber, S. M., \& Gonzalez, J. J. 1992, ApJ, 398, 69

\ref
Zinn, R., \& West, M. 1984, ApJS, 55, 45

\ref
Zinnecker, H., Cannon, R., Hawarden, T., \& MacGillivray, H. 1985, in
The Virgo Cluster of Galaxies, edited by O. Richter \& B. Binggeli
(European Southern Observatory, Garching), p.135

\vfill\break


\parindent=0pt
\centerline{FIGURE CAPTIONS}
\bigskip

{\it Figure 1}.
Flux-calibrated spectra of Fornax nucleated dE's and two Galactic
globular clusters, arranged roughly in order of increasing line strength.
The ordinate flux scale was adjusted to approximately reproduce the
number of e$^-$ at 4800 \AA\ in the original co-added spectra (units are
100 e$^-$). Error bars in the upper left corner of each panel give
1 $\sigma$ errors estimated from photon statistics in object+sky
spectra.

\bigskip
{\it Figure 2}.
Relationship between indices related to metallicity, for Fornax dwarfs
({\it squares}) and two globular clusters ({\it starred symbols}). The
stronger-lined globular is 47~Tuc. Error bars were computed from photon
statistics only. The errors on FCC~150 represent typical uncertainties.
Errors are also shown for the 3 dwarfs with poorly defined indices. The
regression lines were computed assuming errors on both variables (see
text for details).

\bigskip
{\it Figure 3}.
The mean H$\beta$ index is plotted against metallicity derived by us for
dE's ({\it squares}) and two globular clusters ({\it crosses}). The data
point for FCC~243 represents the visually estimated metallicity. The
line is the fit to globular cluster of Brodie \& Huchra (1990). It
represents the locus of old stellar populations of different abundances.
Error bars were computed from count statistics.

\bigskip
{\it Figure 4}.
The equivalent width of H$\beta$ vs $U - B$ colors. The squares with
error bars denote Fornax dE's data in Table 4. Colors are from Caldwell
\& Bothun (1987). The dots represent globular clusters which have
H$\beta$ measured by Brodie \& Huchra (1990), and (dereddened) colors
from Reed \etal (1988).

\bigskip
{\it Figure 5}.
Correlation between spectroscopically derived [Fe/H] values and $U-B$
($a$) and $B-V$ ($b$), for Fornax dwarfs ({\it squares}) and 47~Tuc,
NGC~2808 ({\it crosses}). Error bars as in Figure 3.

\bigskip
{\it Figure 6}.
The $J - K$ color is plotted against the equivalent width of \CAII K for
a small sample of Virgo dE's having both line strengths from Bothun \&
Mould (1988) and infrared colors from Bothun \etal (1985) ({\it filled
squares}) and Thuan (1985) ({\it open squares}). Color uncertainties in
the data are of the order 0.05 mag. Note the magnitude scale offset
between different authors. {\it Dots} connected by line segments refer
to Galactic globular cluster data (\CAII line strength from Da Costa \&
Mould 1988, infrared colors from Brodie \& Huchra 1990).

\bigskip
{\it Figure 7}.
Metallicity is plotted against mean surface brightness in B inside the
effective radius ($<\mu_{\rm B}>_{eff}$), computed from Ferguson's
(1989) data.

\bigskip
{\it Figure 8}.
Contours of galaxy surface density in the Fornax cluster from the
Ferguson \& Sandage (1988)
dwarf sample. The contour unit is one galaxy per $[20^\prime]^2$
Superposed on the contours are the locations of galaxies brighter
than 15th magnitude with redshift less than 2520 \kms and within 4$^\circ$
of NGC 1399. The bar shown on the figure indicates one degree.
North is at the top, east to the right.

\bigskip
{\it Figure 9}.
A slice of redshift space with Right Ascension the azimuthal coordinate.
The radius vectors are 2500 \kms long and ten degrees of RA apart.

\bigskip
{\it Figure 10}.
Redshift distribution of Fornax galaxies: ($a$) dwarf sample, ($b$)
bright sample.


\vfill\break

\bye